\documentclass[prl,twocolumn,superscriptaddress,floatfix,preprintnumbers,amssymb,amsmath]{revtex4}
\usepackage{graphicx}
\usepackage{dcolumn}
\usepackage{bm}
\usepackage[latin1]{inputenc}
\usepackage[mathscr]{eucal}
\usepackage{epsfig}
\usepackage{rotating}

\begin{document}
\preprint{ }

\title{Hybrid single-electron transistor as a source of quantized electric current}

\author{Jukka P. Pekola}
\affiliation{Low Temperature Laboratory, Helsinki University of
Technology, P.O. Box 3500, 02015 TKK, Finland}
\author{Juha J. Vartiainen}
\affiliation{Low Temperature Laboratory, Helsinki University of
Technology, P.O. Box 3500, 02015 TKK, Finland}
\author{Mikko M\"ott\"onen}
\affiliation{Low Temperature Laboratory, Helsinki University of
Technology, P.O. Box 3500, 02015 TKK, Finland}
\affiliation{Laboratory of Physics, Helsinki University of
Technology, P.O. Box 4100, 02015 TKK, Finland}
\author{Olli-Pentti Saira}
\affiliation{Low Temperature Laboratory, Helsinki University of
Technology, P.O. Box 3500, 02015 TKK, Finland}
\author{Matthias Meschke}
\affiliation{Low Temperature Laboratory, Helsinki University of
Technology, P.O. Box 3500, 02015 TKK, Finland}
\author{Dmitri V. Averin}
\affiliation{Department of Physics and Astronomy, Stony Brook
University, SUNY, Stony Brook, NY 11794-3800, USA}

\maketitle
{\bf The basis of synchronous manipulation of individual electrons
in solid-state devices was laid by the rise of single-electronics
about two decades ago~\cite{averin91,SET92,devoret92}. Ultra-small
structures in a low temperature environment form an ideal domain of
addressing electrons one by one. A long-standing challenge in this
field has been the realization of a source of electric current that
is accurately related to the operation frequency $f$
\cite{averin91}. There is an urgent call for a quantum standard of
electric current and for the so-called metrological triangle, where
voltage from Josephson effect and resistance from quantum Hall
effect are tested against current via Ohm's law for a consistency
check of the fundamental constants of Nature, $\hbar$ and $e$
\cite{piquemal04}. Several attempts to create a metrological current
source that would comply with the demanding criteria of extreme
accuracy, high yield, and implementation with not too many control
parameters have been reported. However, no satisfactory solution
exists as yet despite many ingenious achievements that have been
witnessed over the years
\cite{geerligs90,pothier92,keller96,shilton96,fujiwara04,delsing05,vartiainen06}.
Here we propose and prove the unexpected concept of a hybrid
metal-superconductor turnstile in the form of a one-island
single-electron transistor with one gate, which demonstrates robust
current plateaus at multiple levels of $ef$ within the uncertainty
of our current measurement. Our theoretical estimates show that the
errors of the present system can be efficiently suppressed by
further optimizations of design and proper choice of the device
parameters and therefore we expect it to eventually meet the
stringent specifications of quantum metrology.}
\begin{figure}[t!]
\includegraphics[width=1.0\columnwidth,clip]{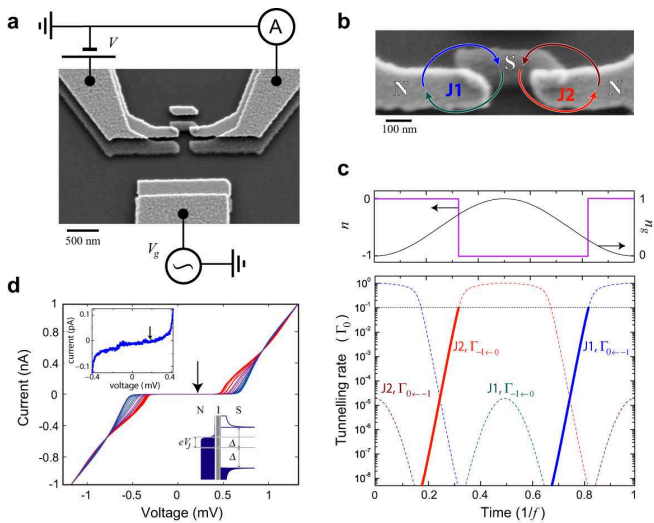}
\caption{The hybrid turnstile and its basic characteristics. {\bf
a}, An electron micrograph of the measured NSN turnstile. It is a
single-electron transistor fabricated by standard electron beam
lithography. The leads are made of copper metal (N) and the island,
the small grain in the centre, is superconducting aluminium (S). The
measurement configuration is added to this image: DC bias voltage
$V$ is applied across the transistor, and a voltage composing of DC
and AC components acts on the gate electrode. {\bf b}, A magnified
image of the island, indicating notation in {\bf c}. {\bf c}, A
basic pumping cycle of the turnstile. The normalized gate voltage
$n_g \equiv C_gV_g/e$, and the instantaneous charge number $n$ on
the island are shown in the top panel against time over one period.
In the bottom frame we show the relevant tunnelling rates, in units
of $\Gamma_0 \equiv \Delta/(e^2R_T)$ through junctions J1 (left one)
and J2 (right one), respectively. Besides the dominant forward
processes, the two most important backward rates are shown. The
tunnelling occurs when $\Gamma$ is of order frequency $f$. Note that
when it takes place for instance through junction J2 in the charge
state $n=0$, the island transits into $n=-1$ state, and the system
stays in this state for a while because all the tunnelling rates for
$n=-1$ state are vanishingly small right after this event. In one
full cycle one electron is transferred through the turnstile from
left to right. {\bf d}, Current-voltage ($IV$) characteristics
measured at various values of DC gate voltage with no AC voltage
applied. The separation of the extreme $IV$ curves is a signature of
the charging energy of the device. The arrow marks the working point
in the turnstile experiments unless otherwise stated. The top inset
shows a magnification of the $IV$ within the gap region
demonstrating high sub-gap resistance of above 10 G$\Omega$. The
lower inset depicts the energy diagram for one junction biased at a
voltage $V_J$. Normal metal is to the left of the barrier in the
centre, and the superconductor to the right, with forbidden states
within the energy interval $2\Delta$ around the Fermi level.}
\label{fig:fig1}
\end{figure}

Synchronized sources, where current $I$ is related to frequency by
$I=Nef$ and $N$ is the integer number of electrons injected in one
period, are the prime candidates for the devices to define ampere in
quantum metrology. The accuracy of these devices is based on the
discreteness of the electron charge and the high accuracy of
frequency determined from atomic clocks. Modern methods are
replacing classical definitions of electrical quantities; voltage
can be derived based on the AC Josephson effect of
superconductivity~\cite{shapiro63} and resistance by quantum Hall
effect~\cite{klitzing80,paalanen82}, but one ampere still needs to
be determined via the mutual force exerted by leads carrying the
current. Early proposals of current pumps for quantum metrology were
based on arrays of mesoscopic metallic tunnel
junctions~\cite{geerligs90,pothier92}, in which small currents could
eventually be pumped at very low error rates~\cite{keller96}.
However, these multijunction devices are hard to control and
relatively slow~\cite{zimmerman03}. Thus, the quest for feasible
implementation with a possibility of parallel architecture for
higher yield have lead to alternative solutions such as
surface-acoustic wave driven one-dimensional channels
\cite{shilton96}, superconducting devices
\cite{niskanen03,zorin04,governale05,kopnin06,mooij06,vartiainen06,cholascinski07},
and semiconducting quantum dots~\cite{ritchie07}. These do produce
large currents in the nano-ampere range but their potential accuracy
is still limited.

Somewhat surprisingly, a simple hybrid single-electron transistor,
with normal metal (N) leads and a small superconducting (S) island,
see Fig. \ref{fig:fig1}, has been overlooked in this context.
As it has turned out in the present work, an SNS-transistor, or
alternatively an NSN-transistor, presents ideally a robust turnstile
for electrons showing accurately positioned current plateaus. We
emphasize here that a one island turnstile does not work even in
principle without the hybrid design. An important feature in the
present system is that hybrid tunnel junctions forbid tunnelling in
an energy range determined by the gap $\Delta$ in the density of
states of the superconductor, see Fig. \ref{fig:fig1}d inset;
current through a junction vanishes as long as $|V_J| \lesssim
\Delta/e$.

Figure \ref{fig:fig1}a shows the simple electric configuration to
operate a hybrid turnstile. A DC bias voltage $V$ is applied between
the source and drain of the transistor and a voltage $V_g$ with DC
and AC components at the gate. To understand the operation of the
turnstile on a more quantitative level, let us follow a basic
operation cycle shown in Fig. \ref{fig:fig1}c. In general, a
sinusoidal AC gate voltage is superposed on the DC offset such that
the total instantaneous voltage on the gate, normalized into charge
in units of $e$, reads $n_g \equiv C_g V_g/e = n_{g0}+A_g \sin(2\pi
f t)$ at frequency $f$. Here $C_g$ is the capacitance of the gate
electrode to the transistor island. In Fig. \ref{fig:fig1}c we have
chosen the gate offset $n_g$ and amplitude $A_g$ to be
$n_{g0}=A_g=0.5$, and the bias voltage across the transistor is set
at $V=\Delta/e$ to suppress tunnelling errors, as will be discussed
below. The key point in the operation of the hybrid turnstile is
that the charge state locks to a fixed value in any part of the
operational cycle except at the moment when a desired tunnelling
event occurs. This is the key feature of the device, originating
from the interplay of the superconducting gap in the energy spectrum
and Coulomb blockade of single-electron tunnelling. It renders this
structure to work as an accurate turnstile where errors can be
suppressed efficiently by decreasing temperature and by choosing the
bias point properly within the superconducting gap. This locking
mechanism is illustrated and explained in Fig.~\ref{fig:fig1}c for
one operational cycle. On the contrary, in the biased NNN transistor
with Coulomb blockade alone, non-synchronized almost frequency
independent DC current through the device is observed~\cite{SET92}.
Likewise, a corresponding fully superconducting SSS device is not
favourable either, because inevitable supercurrent of Cooper pairs
induces significant leakage errors~\cite{zorin03}.

Several hybrid turnstiles with aluminium as the superconductor,
copper as the normal metal, and aluminium oxide as the tunnel
barrier in between were fabricated by standard electron beam
lithography. Both the aluminium and the copper films were 50 nm
thick. Figure \ref{fig:fig1}a shows the NSN sample whose data we
present here. The charging energy of the aluminium island, $E_C
=e^2/(2C_\Sigma)$, is $E_C/k_B \simeq 2$ K, where $C_\Sigma$ is the
total capacitance.
The sum of the tunnel resistances of the two junctions is 700
k$\Omega$, i.e., 350 k$\Omega$ per junction on the average. The
current-voltage ($IV$) characteristics of the transistor are shown
in Fig.~\ref{fig:fig1}d at various values of the DC gate voltage and
with no AC gate voltage applied. The superconducting gap suppresses
the current strongly in the bias region $|V| \lesssim 0.4$ mV.
Outside this region the typical gate modulation pattern shows up
\cite{SET92}. The charging energy of the device was determined based
on the envelopes of these $IV$ curves.

\begin{figure}[t!]
\includegraphics[width=\columnwidth,clip]{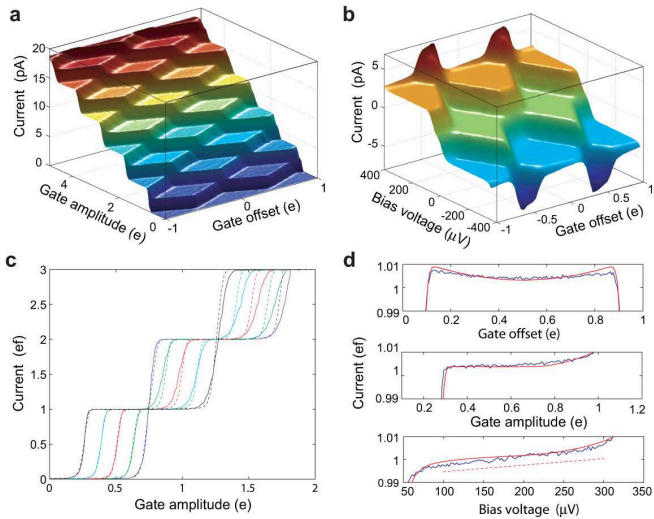}
\caption{Measured characteristics of the NSN turnstile at $f=20$
MHz. {\bf a}, Current plateaus $I=Nef$ as a function of DC gate
offset and AC gate amplitude. The diamond structure is shown up to
$N=10$ steps here. {\bf b}, The plateaus as a function of the bias
voltage $V$ across the transistor up to the gap threshold of about
400 $\mu$V, cf. Fig.~\ref{fig:fig1}d. Current is shown here in a 3D
plot at different DC gate offset positions with a constant gate
amplitude $A_g=0.5$. {\bf c} Current at $V = 200$ $\mu$V measured at
various values of DC gate offset and as a function of gate amplitude
$A_g$ (solid lines). {\bf d}, The $N=1$ plateau measured around the
centre of each diamond in directions of DC gate offset, bias voltage
$V$ across the turnstile and AC gate amplitude, respectively, from
top to bottom. The dashed lines in {\bf c} and the red solid lines
in {\bf d} show the theoretical results according to the sequential
tunnelling model. Here we have used the parameter values $R_T=350$
k$\Omega$ and $E_C/k_B = 2$ K from the DC $IV$ curves (Fig.
\ref{fig:fig1}d), and electron temperature of 80 mK. We further used
$\Delta = 185$ $\mu$eV and sub-gap leakage of $2.5\cdot 10^{-4}$ of
the asymptotic resistance. The dashed line in the bottom panel of
{\bf d} has a slope of 10 G$\Omega$, suggesting that the measured
slope here and in that of the DC $IV$ curve in Fig.~\ref{fig:fig1}d
have the same origin. All the measured currents in the paper have
been multiplied by the same factor 1.004 for the best consistency
with the model: this is well within the $\pm 2$ \% calibration
accuracy of the gain of the current pre-amplifier used.}
\label{fig:fig2}
\end{figure}

The turnstile experiments were carried out by voltage biasing the
transistor at $V\simeq \Delta/e$, highlighted by an arrow in Fig.
\ref{fig:fig1}d. Figure \ref{fig:fig2} shows the current through the
NSN turnstile under varying parameters $n_{g0},A_g$, and $V$ at a
fixed frequency $f=20$ MHz. Figure \ref{fig:fig2}c shows
cross-sections of the 3D plot in Fig.~\ref{fig:fig2}a along
different constant values of $n_{g0}$ against the gate amplitude
$A_g$. The corresponding prediction based on sequential tunnelling
theory~\cite{averin91} is shown by the dashed lines in the same
plot. The experimental data follow the theoretical prediction very
closely. Moreover, the wide flat plateaus at $I=Nef$ seem indeed
promising for metrological purposes. The magnitude of the pumped
current is robust against fluctuations in relevant parameters. It is
not sensitive to exact dimensions or symmetry of the device,
operational temperature, gate offset or its amplitude, or the exact
form of the driving signal in general. Some of these dependencies
are demonstrated in Fig.~\ref{fig:fig2}d based on our present
measurements.

\begin{figure}[t!]
\includegraphics[width=\columnwidth,clip]{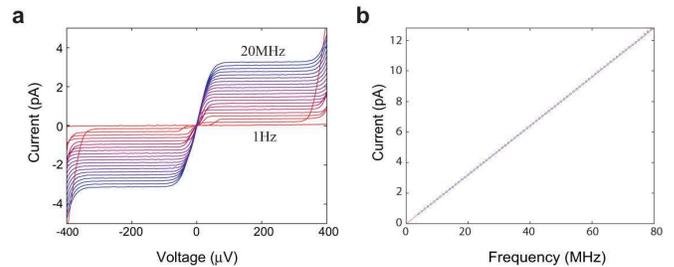}
\caption{The frequency dependence of the NSN turnstile operation.
{\bf a}, $IV$ curves measured at different frequencies ranging from
0 to 20 MHz, at gate settings corresponding to the centre of the
first ($N=1$) current plateau. {\bf b}, The measured current at the
centre of the $N=1$ plateau at the fixed bias of $V=200$ $\mu$V.
Linear dependence up to 80 MHz corresponding to $I\simeq 13$ pA can
be seen.
} \label{fig:fig3}
\end{figure}

Figure \ref{fig:fig3} illustrates the frequency dependence of the
NSN turnstile.
In Fig.~\ref{fig:fig3}a we show the $IV$ curves measured for
$n_{g0}\simeq A_g\simeq 0.5$ at various frequencies. The frequency
dependence of the current corresponding to the first plateau in
measurements of the type that were shown in Fig.~\ref{fig:fig2}b is
plotted in Fig.~\ref{fig:fig3}b against $ef$ in the frequency regime
up to 80 MHz. The predicted $I=ef$ relation is followed closely
within smaller than 1\% deviations in absolute current throughout
this range. We stress here that in the present measurement, using
just a room temperature current amplifier, we cannot test the
agreement between the prediction and the absolute value of the
measured current to any higher degree than this.

Next we discuss the choice of the operating conditions of a hybrid
turnstile and the potential accuracy of this device. Within the
classical model of sequential single-electron tunnelling, the bias
voltage $V$ across the turnstile is a trade-off: small bias leads to
tunnelling events in the backward direction and large $V$ to errors
due to replacement of the tunnelled charge by another one tunnelling
in the forward direction through the other junction. Unwanted events
of the first type occur at the relative rate of $\simeq
\exp(-\frac{eV}{k_BT_N})$, where $T_N$ is the temperature of the
normal metal electrodes. Errors of the second type occur at the
relative rate of $\sim \exp(-\frac{2\Delta -eV}{k_BT_N})$. The
prefactor of this expression is of the order of unity in relevant
cases of interest. Minimizing these errors thus yields $eV \simeq
\Delta$, which is chosen as the operation point in the experiments.
At this bias point the two errors are of order
$\exp(-\frac{\Delta}{k_BT_N})$. For $\Delta \simeq 200$ $\mu$eV
(aluminium) and $T_N < 100$ mK that is a standard range of operation
temperature, we obtain an error rate of $\sim 10^{-10}$, which is
sufficiently small as compared to the requested $\sim 10^{-8}$
accuracy of the metrological source~\cite{piquemal04}.

The analysis above neglects several types of errors. High operation
frequency is one source of error: it leads to missed tunnelling
events and to enhanced tunnelling in the wrong direction. These
errors are suppressed approximately as $\exp(-\frac{\Delta}{2\pi f
e^2 R_T})$. For typical parameters, $\Delta$ for aluminium and $R_T
= 50$ k$\Omega$, we then request $f\ll 4$ GHz for accurate
operation. Such a small value of $R_T$ seems acceptable because of
sufficiently strong suppression of co-tunnelling effects in this
system as will be discussed below. With exponential suppression of
errors in $f$, the metrological accuracy limits then the frequency
to $\sim 100$ MHz for turnstiles with aluminium as the
superconductor.
A possible way to increase the speed of the device is to use niobium
as the superconductor, with almost an order of magnitude larger gap.
With ultrasmall junctions, to keep $E_C\sim \Delta$, which is
another criterion to satisfy in order not to miss any tunnelling
events at the chosen bias point, one would be able to increase the
frequency, and the synchronized current, by the same order of
magnitude. This is an attractive yet unexplored possibility.
Additional improvement, about factor of 3 increase in current, could
possibly be achieved by shaping the ac gate voltage to have
rectangular waveform.

Another source of potential errors is the co-tunnelling
\cite{averin90}, i.e., higher order quantum tunnelling processes,
which are limiting the use of short arrays in normal-metal-based
devices~\cite{martinis92}. In a hybrid turnstile, the lowest order
quasiparticle co-tunnelling errors are, however, suppressed ideally
to zero within the superconducting gap as has been shown for a fully
superconducting case in the past~\cite{averin97}. Another process of
the same order which is not suppressed by the superconducting gap is
the tunnelling of Cooper pairs, also called Andreev reflection. In
junctions without pinholes in the barriers, the rate $\Gamma_A$ of
this tunnelling should be quite small, $\Gamma_A/\Gamma_0 \sim
\hbar/\mathcal{N} e^2 R_T$, where the effective number of the
transport modes in the junction of area $A$ can be estimated as
$\mathcal{N}/A \simeq 10^7$ $\mu$m$^2$~\cite{joyez98}. If the
charging energy is large, $E_C>\Delta$, it suppresses direct
tunnelling of Cooper pairs. Analysis shows that the lowest-order
process that limits the accuracy of the hybrid turnstiles with ideal
superconducting electrodes is then the co-tunnelling of one electron
and one Cooper pair, the rate $\Gamma_{CPE}$ of which can be
estimated roughly as $\Gamma_{CPE}/\Gamma_0 \sim
(1/\mathcal{N})(\hbar/e^2 R_T)^2$. These processes are suppressed to
the level mandated by the metrological accuracy,
but one may need to increase the junction resistance to $R_T > 50$
k$\Omega$. Further compromise in the operation frequency would not
be necessary even in this case if one benefits from the improvements
in the turnstile operation discussed above.

Sub-gap leakage, due, for instance, to non-zero density of
quasiparticle states within the gap, introduces a material- and
fabrication-specific source of errors into our system. This effect
is demonstrated by the non-vanishing slope of the $IV$ curve in the
top inset of Fig.~\ref{fig:fig1}d and by an equal slope in the bias
dependence of the bottom panel in Fig.~\ref{fig:fig2}d showing the
current on the first plateau. Our estimates show that as far as
co-tunnelling is concerned, such errors are small already in the
present devices. For sequential tunnelling sub-gap leakage causes a
substantial additional contribution to current, of order $10^{-3}$
in the present device. With high quality tunnel junctions, possibly
by an improved fabrication process, its influence can be suppressed
further. Furthermore, the separation of the current plateaus, with
one bias polarity only, is ideally not sensitive to this effect,
unlike the absolute value of current on a single plateau. Yet we
find the sub-gap leakage as the main issue to be solved in order to
realize a metrologically compatible turnstile. We would also like to
point out that a series connection of a few SN junctions would
present an improved version of a multijunction electron
pump~\cite{pothier92,keller96} in terms of leakage and co-tunnelling
errors, since this device can be operated without external bias
voltage.

There is a difference between the performance of the SNS and NSN
structures, which was not emphasized in the discussion above. The
charge transport in these systems is associated with non-trivial
heat flux: based on the same strategy as discussed here a
single-electron refrigerator can be realized
\cite{pekola07,saira07}. In this device superconductor is always
heated, but under proper bias conditions heat flows out from the
normal metal. Therefore, in a single island realization, an SNS
configuration is more favourable, at least theoretically, as
compared to the NSN turnstile. It turns out that with quite
realistic parameters it is possible to refrigerate the small,
thermally well isolated N island of an SNS turnstile substantially,
and hence the error rates can be further suppressed. The source and
drain leads can be thermalized close to the bath temperature by
proper choice of geometry and materials.
Cooling or overheating effects were, however, not identified
definitely in the present experiment, where small errors ($\ll
10^{-3}$) could not be assessed.

One of the key advantages of the single-island turnstile, as
compared to multi-island pumps is that the influence of the
background charges~\cite{backgroundcharges} can be compensated by
adjusting just a single DC gate voltage, e.g., to maximize the width
or minimize the slope of the current plateaus. Therefore the level
of the current can be increased by a relatively straightforward
parallelization of several turnstiles.
If an upgrade of current by, for example, an order of magnitude is
necessary, DC-gate setting of each of the ten turnstiles can be
adjusted individually whereafter the currents of them can be
combined. The whole device can then be operated with common-to-all
DC bias and AC gate drive.

{\bf Acknowledgements} We thank Mikko Paalanen and Antti Manninen
for fruitful discussions, and Antti Kemppinen for assistance in the
measurements. The work was financially supported by Technology
Industries of Finland Centennial Foundation and by Academy of
Finland. M. M. acknowledges
the Finnish Cultural Foundation for financial support.\\ \\



\end{document}